\documentclass[12pt,a4paper]{article}

\usepackage{amsmath,amsfonts,latexsym,amssymb,amsthm}
\usepackage{verbatim,amsthm,curves,graphics}
\usepackage{mathrsfs}

\begin{document}

\theoremstyle{definition}
\newtheorem{defn}{Definition}[section]
\newtheorem{remark}{Remark}[section]
\newtheorem{thm}{Theorem}[section]
\newtheorem{lem}{Lemma}[section]
\newtheorem{prop}{Proposition}[section]
\newtheorem{cor}{Corrollary}[section]
\renewcommand{\thesection}{\Roman{section}}
\newcommand{\tr}{\operatorname{tr}}
\newcommand{\vol}{{\rm vol}}
\newcommand{\g}{{g}}
\renewcommand{\i}{{\rm i}}
\newcommand{\supp}{\operatorname{supp}}
\newcommand{\sign}{\operatorname{sign}}
\newcommand{\id}{\operatorname{id}}
\newcommand{\cO}{\mathcal{O}} 
\newcommand{\cC}{\mathscr{C}} 
\newcommand{\cE}{\mathscr{E}} 
\newcommand{\cD}{\mathcal{D}} 
\newcommand{\cA}{\frak{A}} 
\newcommand{\cK}{\mathscr{K}} 
\newcommand{\cH}{\mathscr{H}} 
\newcommand{\cV}{\mathscr{V}} 
\newcommand{\cI}{\mathscr{I}} 
\newcommand{\cF}{\mathscr{F}} 
\newcommand{\cN}{\mathscr{N}} 
\newcommand{\cX}{\mathscr{X}} 
\newcommand{\cY}{\mathscr{Y}} 
\newcommand{\mn}{\Bbb{N}} 

\newcommand{\mc}{\Bbb{C}} 

\newcommand{\mi}{\Bbb{I}} 

\newcommand{\mj}{\Bbb{J}} 

\newcommand{\mr}{\Bbb{R}} 

\newcommand{\mslash}{/\!\!\!}             
\newcommand{\slom}{/\!\!\!G}         
\newcommand{\dirac}{/\!\!\!\nabla}        
\newcommand{\myid}{\leavevmode\hbox{\rm\small1\kern-3.8pt\normalsize1}}
\newcommand{\rso}{|\!|\!|}
\newcommand{\lso}{|\!|\!|}

\title{The State Space of Perturbative Quantum Field Theory in 
Curved Spacetimes}
\author{Stefan Hollands\thanks{Electronic mail: 
        \tt stefan@bert.uchicago.edu}\\    		
        \it{Department of Physics, Enrico Fermi Institute,}\\
	\it{University of Chicago, 
            5640 Ellis Ave.,} \\ 
        \it{Chicago, IL 60367, U.S.A.}
\and
	Weihua Ruan\thanks{Electronic mail: \tt ruanw@calumet.purdue.edu}\\
	\it{Department of Mathematics, Computer Science and Statistics,}\\
	\it{Purdue University Calumet,}\\
 	\it{Hammond, IN 46323, U.S.A.}
}
\date{\today}

\maketitle
\begin{abstract}
The space of continuous states of perturbative interacting
quantum field theories in globally hyperbolic 
curved spacetimes is determined. Following 
Brunetti and Fredenhagen, we first define an abstract algebra of observables 
which contains the Wick-polynomials of the free field as well as 
their time-ordered products, and hence, by the well-known 
rules of perturbative quantum field theory, 
also the observables (up to finite order) of interest for the 
interacting quantum field theory. We then determine the space of continuous  
states on this algebra. Our result is that this space 
consists precisely of those states whose truncated $n$-point functions
of the free field are smooth for all $n \neq 2$, and whose
two-point function has the singularity structure of a Hadamard 
fundamental form. 
A crucial role in our analysis is played by the positivity 
property of states. On the technical side, 
our proof involves functional analytic methods, in 
particular the methods of microlocal analysis. 

\end{abstract}

\section{Introduction}

The perturbative construction of self-interacting quantum field theories in 
Minkowski spacetime was put on a completely rigorous mathematical 
footing in the works by Bogliubov, Parasiuk, Hepp, Zimmermann 
and other people \cite{bphz} in the late sixties and early seventies. 
The issue of generalizing these constructions to curved spacetimes 
was first analyzed by Bunch and collaborators \cite{bu, p}. These 
authors showed, within the context of Euclidean quantum field theory on 
Riemannian curved spaces, that if a theory is 
``perturbatively renormalizable'' 
in flat space, then it remains so in curved space. However, while 
the perturbative definition of a quantum field theory 
on flat Euclidean space gives rise, via a ``Wick rotation'', to 
the definition of a corresponding theory on Minkowski space, 
no such connection holds for curved Lorentzian spacetimes, which, 
apart from a few special classes of spacetimes such as static ones,
do not possess a corresponding real Riemannian section. This means
that Euclidean methods cannot directly be used for the definition 
of interacting quantum field theories in most Lorentzian spacetimes.

Significant progress in perturbative construction of interacting 
quantum field theories on an arbitrary 
globally hyperbolic Lorentzian spacetime has recently been made by 
\cite{bfk, bf, hw}, using the mathematical tools of ``microlocal
analysis'' \cite{h}. In \cite{bfk}, the authors demonstrated that the formally
infinite Wick-polynomials of a free field can be given a well-defined
sense as operator-valued distributions via a normal ordering prescription. 
In \cite{bf}, they then constructed time-ordered products of these
Wick-polynomials. As in Minkowski spacetime, some 
``renormalization ambiguities'' necessarily arise in the 
definition of the time-ordered 
products in curved spacetime, and moreover, unlike in Minkowski space, 
renormalization ambiguities also arise in the definition of the 
Wick-polynomials in curved spacetime. 
If one demands that these 
quantities be locally constructed from the metric in a
generally covariant way, have a certain
scaling behavior under a rescalings of the metric 
and have a suitable dependence under variations of 
the metric, then it can be shown \cite{hw}, that these 
renormalization ambiguities are reduced 
to a finite number of free parameters.\footnote{
It turns out \cite{hw} that the normal 
ordered Wick-polynomials and their time-ordered products defined
in \cite{bfk, bf} necessarily fail to be locally constructed 
out of the metric in a covariant manner. A construction of 
Wick-polynomials that are locally defined in terms of the metric
in a covariant manner and have the above additional 
properties was given in \cite{hw}. Work on a corresponding
construction for the time-ordered products is in progress \cite{hw1}.} 
Moreover, a detailed analysis of 
the nature of these renormalization ambiguities \cite{hw} 
leads to the conclusion that interacting quantum 
field theories in globally 
hyperbolic spacetimes have the same classification 
into ones that are perturbatively 
renormalizable and ones that are not as in Minkowski space.

The (smeared) Wick-polynomials and their (smeared)
time-ordered products may be regarded as members of some  
abstract *-algebra, $\mathcal W$.\footnote{
While the construction of this algebra involves
the choice of a quasi-free Hadamard of the corresponding free field theory, 
it turns out \cite{hw} that different choices for this state
give rise to isomorphic algebras. Therefore, as an abstract 
*-algebra, $\mathcal W$ is independent of that choice.}
The Wick-polynomials and time-ordered
products in $\mathcal W$ which satisfy the above 
additional locality, covariance and scaling requirements 
can be used to define, via the usual perturbation 
expansions familiar from Minkowski space, the quantities of 
interest in the interacting theory. 
The infinite sums occurring in these perturbation expansions do
not by themselves define elements of the algebra $\mathcal W$. However, 
if these sums are truncated at some arbitrary finte order, then the 
so obtained truncated expressions {\it are} elements of 
the algebra $\mathcal W$. This algebra therefore contains 
all observables of interest in the interacting theory up 
to an arbitrary finite order in perturbation 
theory. 

\medskip

In this work we investigate the space of quantum states on 
$\mathcal W$, that is, the space of states for the perturbatively 
defined interacting quantum field theory. Here, by a state 
we mean a linear functional $\psi: \mathcal W \to \mc$ 
which is normalized so that $\psi(I) = 1$, where 
``$I$'' denotes the identity element in $\mathcal W$, and which is 
positive in the sense that $\psi(A^*A) \ge 0$ for any element
$A$ in $\mathcal W$. The above algebraic notion of state is 
related to the usual Hilbert space notion of state, but it 
is more general: Given a representation of $\mathcal W$ on 
some Hilbert space, one can consider a vector or density matrix 
state\footnote{Actually, we must restrict ourselves here to the 
vector or density matrix states contained in some common, dense invariant 
domain.}
in this Hilbert space as defining a corresponding algebraic state on 
$\mathcal W$. However, it is well-known that 
not all algebraic states---and not even all physically 
interesting ones---can be obtained in this 
manner from some specific Hilbert space representation\footnote{
For example, the standard thermal state at some finite 
temperature of the free field on Minkowski space gives rise 
to a state on $\mathcal W$. But this state cannot be regarded
in any sense as arising from a density matrix state in the vacuum 
representation.}. 

It was conjectured in \cite{hw} that the space of states on $\mathcal
W$ consists precisely of those positive, normalized linear functionals 
$\psi$ whose truncated $n$-point functions of the free field are 
smooth for $n \neq 2$ and whose two-point function of the free
field is of Hadamard form. 

The main result of our paper (Thm.~\ref{mainthm})
is that this conjecture is correct with 
regard to the states $\psi$ on $\mathcal W$ that 
are continuous with respect to some natural topology on the algebra
$\mathcal W$. In order to clarify the status of this continuity 
requirement, we note 
that, if no restriction at all was placed on the state $\psi$, then its 
$n$-point functions of the free field would be merely linear functionals
on the space of testfunctions, but would not even have to be distributions, 
which is always assumed even in such general frameworks as 
the Wightman-axioms. On the other hand, the continuity requirement 
that we shall impose on the states under consideration will automatically
imply that these $n$-point functions are at least distributions.
In addition, since any element of $\mathcal W$ can be obtained as a limit of 
elements in the subalgebra $\mathcal A$ spanned by finite 
products of (smeared) free fields, a continuous state on $\mathcal W$ 
is uniquely determined by its restriction to 
the subalgebra $\mathcal A$, that is, by its $n$-point functions of the 
free field. This is in complete agreement with the philosophy behind the 
so-called ``point-splitting'' prescription for renormalizing 
Wick-products such as the stress energy operator---which is an element of 
$\mathcal W$, but not of $\mathcal A$---wherein one defines the 
action of a state on a Wick-product as the limit of the expectation
values of suitable ``point-split'' quantities, which are elements 
of $\mathcal A$. Clearly, such a prescription implicitly involves a 
continuity assumption about the action of states on the algebra 
$\mathcal W$, which, as one can show, is a special case of the 
general continuity assumption considered in this paper.

An equivalent way to express our result is to say  
that only those states on the algebra $\mathcal A$ of free fields
have a continuous extension to the algebra 
$\mathcal W$ of observables in perturbation theory 
whose truncated $n$-point functions are smooth for 
$n \neq 2$ and 
whose two-point function is of Hadamard form.
It has long been known from the 
theory of renormalizing the stress energy operator
that there exist many states for the free field whose 
action cannot be extended in a reasonable (that is, 
continuous) way from 
free fields to the stress energy operator. Our result puts 
this observation into a much more general perspective. 

We note that our result does not hold in general for functionals
on $\mathcal W$ which are continuous but which are not
positive: One can construct continuous functionals on $\mathcal W$
whose truncated $n$-point functions are not smooth.

For simplicity and definiteness, we will here consider only the 
case of a Hermitian scalar field. However, the generalization of 
our results to other types of fields should be possible. 

\medskip

The organization of this paper is as follows: We first review 
the definition of the basic algebra of free fields, $\mathcal A$.
After that, we recall the definition of the truncated $n$-point
functions of a state on $\mathcal A$ and of Hadamard states, 
thereby giving a brief summary of some notions from microlocal analysis
that shall be needed later on. We then briefly review the 
construction of the algebra $\mathcal W$ and recall 
how the topology on $\mathcal W$ is defined. After that, 
we present our main result, Thm.~\ref{mainthm}. The proof 
of that result makes up the rest of the paper.
Some parts of this proof are moved to an appendix.

\vspace{2cm}
{\bf Acknowledgements:} We would like to thank R. M. Wald for helpful 
discussions during the early stages of this research. We are indebted
to him in particular for suggesting to us that the positivity property 
of states should play a crucial role in the proof of our main 
result. S. Hollands was supported by NFS grant PHY00-90138 to 
the University of Chicago. 

\pagebreak

\section{Preliminaries}
\subsection{Definition of the minimal algebra $\mathcal A$ of observables for 
the free Klein-Gordon field}

A free classical Hermitian Klein-Gordon field on a curved spacetime 
is a real valued solution of the equation 
\begin{equation}
(\Box - \xi R - m^2) \varphi = 0, 
\end{equation}
where $\Box
= |g|^{-1/2} \partial_\mu |g|^{1/2} g^{\mu\nu} \partial_\nu$ is 
the wave-operator in curved space, $R$ is the curvature scalar, and 
where $m, \xi$ are real parameters. It is known that this 
equation possesses unique advanced and retarded
fundamental solutions on any (time-oriented) globally hyperbolic
spacetime. These are determined by the equations
\begin{equation}
(\Box - \xi R - m^2) \Delta_{\rm adv} = 
(\Box - \xi R - m^2) \Delta_{\rm ret} = \delta, 
\end{equation}
and by the requirement that the support of $\Delta_{\rm adv}$
respectively $\Delta_{\rm ret}$
consists of pairs of points $(x_1, x_2)$ such that $x_2$ is 
in the causal past respectively future of $x_1$.  

\medskip

The theory of a quantized free Klein-Gordon field on globally hyperbolic 
spacetimes \cite{wald,kw} can be described in different ways. For our 
purposes, it is essential to use an algebraic approach. 
In this approach one starts with an abstract *-algebra, $\mathcal{A}$, 
of quantum observables for the free field theory. Several
choices for $\mathcal A$ are possible.\footnote{We note that the 
choice for $\mathcal A$ used in this paper is not the same 
as in \cite{wald,kw}, where the authors work instead with
the algebra generated by exponentiated smeared field operators. 
Such a choice has some advantages, but would not be convenient
for our purposes.} We here take 
$\mathcal A$ to be the *-algebra generated 
by the identity $I$ and the 
smeared field operators $\varphi(f)$, sometimes formally
written as $\int \varphi(x) f(x) \, d\eta$, 
with the following relations: 

\begin{itemize}
\item  Linearity: $f \to 
\varphi(f) \in \mathcal{A}$ is complex linear.

\item  Klein-Gordon: The field operators satisfy the Klein-Gordon
equation in the sense that 
$\varphi(( \Box -\xi R -m^{2}) f) =0$. 

\item  Hermiticity: The quantum field is Hermitian, 
$\varphi(f)^* = \varphi(\bar f)$.
\item  Commutation Relations: 
\begin{equation}
\label{com}
[\varphi(f_1), \varphi(f_2)] = 
i \Delta(f_1, f_2) \cdot I, 
\end{equation}
where $\Delta = \Delta_{\rm adv} - \Delta_{\rm ret}$. 
\end{itemize}
We will subsequently consider more general observables than those contained 
in the algebra $\mathcal A$. We therefore refer to this algebra as 
the ``minimal algebra''.

\subsection{States on $\mathcal A$}

Given a state $\psi$ on $\mathcal A$, 
one defines its ``$n$-point functions'', $\psi_n$, 
as the $n$ times multilinear functionals
on the space of testfunctions given by 
\begin{equation}
\psi_n(f_1, f_2, \dots, f_n) = \psi(\varphi(f_1) \varphi(f_2) 
\dots \varphi(f_n)).
\end{equation}
Every state on $\mathcal A$ is uniquely determined by the collection
of its $n$-point functions.
One also defines the ``truncated $n$-point functions'',
$\psi_n^T$, of a state $\psi$. For the first few $n$, these
are given by 
\begin{eqnarray}
\psi_1^T(f) &=& \psi_1(f), \nonumber\\
\psi_2^T(f_1, f_2) &=& \psi_2(f_1, f_2) - \psi_1(f_1)\psi_1(f_2),
\nonumber\\
\psi_3^T(f_1, f_2, f_3) &=& \psi_3(f_1, f_2, f_3) - 
 \psi_1(f_1)\psi_2(f_2, f_3) -
\nonumber\\ 
&&  \psi_1(f_2)\psi_2(f_1, f_3) - 
\psi_1(f_3)\psi_2(f_1, f_2) + 
2\psi_1(f_1) \psi_1(f_2) \psi_1(f_3). 
\end{eqnarray}
Their definition for general $n$ is as follows. Denote by 
$\mathcal{I}_n$ the set of partitions $P$ of the set $\{1, \dots, n\}$
into pairwise disjoint, ordered subsets $r_1, \dots, r_j$. If 
$r$ is a set in $P$, then we denote its elements by $r(1),\dots, 
r(|r|)$, where $|r|$ is the number of elements in the set $r$. 
Note that by definition $r(i) < r(j)$ if $i < j$. With this 
notation, the truncated $n$-point functions are implicitly defined
by 
\begin{equation}
\label{truncf}
\psi_n(f_1, f_2, \dots, f_n) = 
\sum_{P \in \mathcal{I}_n} \prod_{r \in P}
\psi^T_n(f_{r(1)}, f_{r(2)}, \dots, f_{r(|r|)}).
\end{equation}
Note that the sum always contains the term $\psi_n^T(f_1, \dots, f_n)$ 
corresponding to the trivial partition
consisting only of the set $\{1, \dots, n\}$. 
Therefore, once the truncated $n$-point functions 
have been defined for $1, \dots, n-1$, one can solve the above
relation for $\psi^T_n$ in terms of $\psi_n$ and 
the lower order truncated $n$-point functions. 

\medskip

A state on $\mathcal{A}$ is called ``quasi-free'' if its truncated 
$n$-point functions are all zero except for $n=2$. A standard example
for a quasi-free state is the vacuum state in Minkowski-space, and, 
more generally, all states constructed from some set of ``positive 
frequency solutions'' to the Klein-Gordon equation. It is a consequence
of the definition \eqref{truncf} that the odd $n$-point functions of 
a quasi-free state vanish, and that the even $n$-point functions can 
be expressed solely in terms of the two-point function. These
expressions can be summarized in the formula\footnote{
Actually, expressions like $e^{i\varphi(f)}$ are not 
elements in our algebra $\mathcal A$, since this algebra contains by 
definition only finite sums of products of smeared free fields. 
What is meant by the formula below
(and other similar formulas in the sequel) is the set of 
equalities obtained by expanding both sides of the equation 
in a formal power series and by equating the corresponding terms
in these series.}
\begin{equation}
\psi(e^{i\varphi(f)}) = e^{-\frac{1}{2}\psi_2(f, f)},  
\end{equation}
valid for any quasi-free state $\psi$. 

\medskip

A state $\psi$ on $\mathcal{A}$ is said to be of ``Hadamard form'' if 
its two-point function has no spacelike singularities, and if it
locally can be written in the form
\begin{equation}
\psi_2(x_1, x_2) = U(x_1, x_2) \sigma^{-1}
+ V(x_1, x_2) \ln \sigma + W(x_1, x_2). 
\end{equation} 
Here, $\sigma$ is the signed, squared 
geodesic distance between the points $x_1$ and $x_2$, 
$U$ and $V$ are certain smooth functions defined in terms 
of the metric and the coupling parameters, 
and $W$ is a smooth function depending on the 
state in question. The $\epsilon$-prescription for the singular
terms $\sigma^{-1}$ and $\ln \sigma$ is the same as for the 
usual vacuum two-point function in Minkowski space. 
Strictly speaking, the quantities $U, V$ and $W$ are well defined only 
for spacetimes which are analytic, so the above definition 
of Hadamard states needs to be modified in spacetimes which are 
only smooth. For a discussion of this and a precise formulation 
of the statement that ``there are no spacelike singularities'', 
we refer the reader to \cite{kw}. 
It is an immediate consequence of the definition that if $\psi$ and
$\omega$ are two Hadamard states, then the difference between the 
corresponding two-point functions, $\psi_2 - \omega_2$, is 
smooth.

\medskip

There exists an alternative, equivalent characterization of Hadamard
states in terms of the so-called ``wave front set'' of its associated
two-point function, which plays an important role in this work. 
In order to state what this characterization is,  we first recall
the concept of the wave front set of a distribution. Let $u$ be a 
smooth function on $\mr^n$ with compact support. Then it is known that
the Fourier transform
\footnote{
Our convention for the Fourier transform is 
$\widehat u(k) = \frac{1}{(2\pi)^{n/2}} \int u(x) e^{+ikx} \, d^n x$.}
of $u$ is rapidly decaying, that is, 
for any $N$, there is a constant $C_N$ such that 
\begin{equation}
\label{rapid}
|\widehat u(k)| \le C_N ( 1 + |k| )^{-N} \quad
\text{for all $k \in \mr^n$}. 
\end{equation}
Let now $u$ be a compactly supported {\it distribution}. Then it is 
known that the Fourier transform $\widehat u$ is polynomially 
bounded in $k$. However, it is no 
longer true in general that $\widehat u$ is rapidly decaying in all 
directions. The directions in $k$-space, for which there 
exists no conic neighborhood (that is, a 
neighborhood which is invariant under multiplication by 
positive scalars) such that \eqref{rapid} holds, are called 
``singular directions of $u$'' and are denoted by $\Sigma(u)$. 
Note that $\Sigma(u)$ is by definition a conic set.

Let now $u$ be an arbitrary distribution on an open set $X \subset 
\mr^n$, not necessarily of compact support. Then one can 
define the singular directions of 
$u$ near some point $x$ by localizing $u$ with a
smooth function of compact support for which $\chi(x) \neq 0$, 
by defining $\Sigma_\chi(u) = \Sigma(\chi u)$.
If one now shrinks the support of $\chi$ to the point $x$, then
one obtains the singular directions of $u$ at the point $x$, defined
as 
\begin{equation}
\Sigma_x(u) = \bigcap_{\chi(x) \neq 0} \Sigma_\chi(u)
\end{equation}
The wave front set, ${\rm WF}(u)$, of $u$ is just the union of all nonzero 
singular directions of $u$, 
\begin{equation}
{\rm WF}(u) = \{(x, k) \in X \times (\mr^n \backslash  \{0\})
\mid k \in \Sigma_x(u) \}.
\end{equation}
Note that it follows directly from the definition of the 
wave front set that a distribution $u$ is given by a 
smooth density if and only if ${\rm WF}(u)$ is empty.
It can be demonstrated that the wave front set transforms 
covariantly under a change of coordinates, that is, if $\phi$ is 
a smooth one--to--one map on $X$, then $(\phi(x), k) \in {\rm WF}(u)$
is equivalent to $(x, [D\phi(x)]^t k) \in {\rm WF}(\phi^* u)$, where 
$D\phi = \frac{\partial \phi}{\partial x}$, where $^t$ means 
the transpose of a matrix and where $\phi^* u$ denotes the pull-back
of a distribution, defined by analogy with the 
pull-back of a smooth density. This makes it possible to 
define in an invariant way the wave front set of a 
distribution on a smooth manifold $X$. The
above transformation law then shows that ${\rm WF}(u)$ is 
intrinsically a conic
subset of the cotangent bundle $T^* X$ minus
its zero section. It is common to 
define, for every closed conic set $\Gamma$, the subspace 
$\cD'_\Gamma(X) = \{u \in \cD'(X) \mid {\rm WF}(u) \subset \Gamma\}$ 
of the space $\cD'(X)$ of all distributions on $X$.

Having introduced the wave front set of a distribution, we can now 
state the promised alternative characterization of Hadamard states: 
Namely, a state $\psi$ is Hadamard if the wave front set of its two-point
function has the following form:
\begin{equation}
{\rm WF}(\psi_2) = \{
(x_1, k_1; x_2, -k_2) \in T^*(M \times M) \backslash \{0\} \mid
(x_1, k_1) \sim (x_2, k_2), \,\, k_1 \in (V_+)_{x_1} \}.
\end{equation}
The notation $(x_1, k_1)
\sim (x_2, k_2)$ means that $x_1$ and $x_2$ can be joined by a null-geodesic
and that the covectors $k_1$ and $k_2$ are cotangent and coparallel to that
null-geodesic. $(V_+)_x$ denotes the closed forward lightcone in the 
cotangent space at the point $x$, defined as the set of all 
future directed timelike or null covectors in the cotangent 
space at $x$. The closed backward lightcone, $(V_-)_x$, is defined similarly.
For later purposes, we also set $V_\pm = \cup_{x\in M} (V_\pm)_x$.  

\medskip

For the algebras considered in this paper, there holds
the so-called GNS-theorem, which says that,  
given an algebraic state $\psi$ on the algebra, 
there is a *-representation $\pi_\psi$ of the algebra  on a 
Hilbert space $\mathcal{H}_\psi$ containing 
vector $|\Omega_\psi \rangle$, which is determined, 
up to equivalence, by the relation $\psi(A) = \langle \Omega_\psi 
| \pi_\psi(A) | \Omega_\psi \rangle$, required to hold 
for all algebraic elements $A$. This representation is commonly 
called the ``GNS-representation'' of 
the state $\psi$. For the case of the algebras 
considered in this paper, the GNS-representations corresponding to 
different states are in general inequivalent. 

\subsection{Definition of the extended algebra $\mathcal{W}$ }

In the previous subsections, we have introduced a minimal algebra, 
$\mathcal A$, of observables for a free Klein-Gordon field, and we have 
introduced the notions of quasi-free states and 
of Hadamard states on this algebra. 
The algebra $\mathcal A$ contains the observables corresponding to 
the smeared $n$-point functions of the free field, $A = 
\varphi(f_1) \varphi(f_2) \dots \varphi(f_n)$ (and finite linear 
combinations thereof). If one wants to define a nonlinear quantum
field theory perturbatively 
off the free field theory, one must consider additional 
observables such as (smeared) Wick-polynomials 
of the free field and (smeared) time-ordered products of these fields. 
However, none of these observables are contained in $\mathcal A$. 

In order to include these additional observables, 
we consider, besides the minimal algebra $\mathcal A$, 
an enlarged algebra of observables, $\mathcal W$, that contains 
$\mathcal A$ and that also contains, among others, elements corresponding
to (smeared) Wick-polynomials of free fields and time-ordered products
of these fields. The construction of the algebra $\mathcal{W}$
was first given by \cite{bf} and was later formalized in \cite{fd} 
for the case of Minkowski spacetime. The straightforward generalization 
of \cite{fd} to curved spacetimes can be found in \cite{hw}.
The construction of $\mathcal{W}$ initially depends on the choice of 
some quasi-free Hadamard state $\omega$. One can show however \cite{hw}
that different choices for $\omega$ give rise to isomorphic algebras
$\mathcal W$, so in this sense $\mathcal W$ does not depend on the 
specific choice for $\omega$.  

The observables in the interacting field 
theory (defined perturbatively off the free field theory) are 
given in terms of the well-known formal power series expansions
in the coupling constant, whose coefficients are elements of 
the algebra $\mathcal W$. 
The infinite sums occurring in these expressions are, of course, 
not by themselves elements of $\mathcal W$. 
These series are believed not to converge, and are at best expected to 
approximate the ``true, nonperturbative quantities'' well only 
up to some finite order, after which they diverge. For this reason, 
one is only interested, even in principle, in the calculation of  
the interacting observables up to some finite order in perturbation 
theory anyway. The latter observables {\it are} elements of our algebra 
$\mathcal W$, and we therefore take the view that $\mathcal W$ 
should be regarded as the algebra 
of observables which are of interest in perturbative 
quantum field theory. 

\medskip

For the convenience of the reader, we now recall the basic 
steps in the definition of $\mathcal W$. Let $\omega$ be a
quasi-free Hadamard state on the minimal 
$\mathcal A$, which we shall keep fixed
for the rest of this work. The minimal algebra
$\mathcal A$ contains the normal ordered smeared $n$-point
functions of the free field, defined as\footnote{
Actually, the fields $\varphi(f)$ appearing in the expression below
should be understood as the representers of these algebraic elements
in the GNS-representation of the state $\omega$.
}
\begin{equation}
\label{wickpdef}
:\varphi^{\otimes n}(\otimes_i f_i) :_\omega \,\, \equiv
\,\, :\varphi(f_{1})
\varphi(f_2) 
\ldots \varphi(f_n) :_{\omega } \,\, = 
\frac{\partial^{n}}{i^{n}\partial
t_1 \partial t_2\dots \partial t_n} G(\sum_i t_i f_i)
\Bigg|_{t_1 = t_2 = \dots = 0},
\end{equation}
where 
\begin{equation}
\label{gdef}
G(f) = e^{\frac{1}{2}\omega
_{2}\left( f,f\right) }
e^{i\varphi \left( f\right) }.
\end{equation}
Explicitly, 
\begin{eqnarray}\label{examp}
: \varphi (f) :_\omega \,\, &=& \varphi(f), \nonumber\\
: \varphi (f_1) \varphi(f_2) :_\omega \,\, &=& 
\varphi(f_1)\varphi(f_2) - \omega_2(f_1, f_2) \cdot I, \nonumber\\
: \varphi (f_1) \varphi(f_2) \varphi(f_3) :_\omega \,\, &=& 
\varphi(f_1)\varphi(f_2)\varphi(f_3) - \omega_2(f_1, f_2)\varphi(f_3)-\nonumber\\
&& \omega_2(f_1, f_3)\varphi(f_2)-
\omega_2(f_2, f_3)\varphi(f_1)
\end{eqnarray}
for the first few values of $n$. If $t$ is a smooth 
testfunction on $M^n$, we also define the elements
\begin{equation}
\label{Adef}
A = \,\, :\varphi^{\otimes n}(t) :_\omega
\,\, = \int :\varphi(x_1) \varphi(x_2) \dots \varphi(x_n) :_\omega
\, t(x_1, x_2, \dots, x_n) \,d\eta_1 d\eta_2 \dots d\eta_n 
\end{equation}
in the minimal algebra $\mathcal A$. 

In order to obtain an algebra which is large enough to 
contain the observables of interest in perturbative quantum 
field theory, one would like to smear the fields $:\varphi^{\otimes n}
:_\omega$ not only with smooth test{\it functions} $t$, but also
in addition with certain compactly supported 
test{\it distributions}. Now, smearing the operator-valued distributions
$:\varphi^{\otimes n}:_\omega$ with a distribution involves taking
the pointwise product of two distributions. As it is well-known, 
the pointwise product of two distributions is in general meaningless. 
While it is therefore 
impossible to smear the $:\varphi^{\otimes n}:_\omega$ 
with an arbitrary compactly supported distribution $t$, it turns out 
to be possible to smear it with distributions $t$ contained in a 
subclass $\mathcal{E}^\prime_n$ 
of the class of all compactly supported distributions
(here the Hadamard property of $\omega$ enters). This 
subclass is most conveniently described in terms of the wave front set,  
\begin{multline}\label{edef}
\mathcal{E}_n^\prime 
= \{ \text{symmetric, compactly supported distributions $t$
on $M^n$}\\
\text{with ${\rm WF}(t) \subset T^*M^n \backslash
(V_+^n \cup V_-^n)$} \}.
\end{multline}
\begin{defn}
$\mathcal{W}$ 
is the *-algebra generated by the elements $A$ of the 
form \eqref{Adef} with $t \in \mathcal{E}^\prime_n$. 
\end{defn}
By construction, the extended algebra $\mathcal W$ contains
the minimal algebra $\mathcal A$, but it also contains 
additional elements such as for example normal ordered
Wick-powers at the same spacetime point, 
which are defined as follows: Let 
\begin{equation}
t(x_1, x_2, \dots, x_n) = f(x_1) \delta(x_1, x_2, \dots, x_n), 
\end{equation} 
where $f$ is a compactly supported testfunction.
Then one can show that $t \in \mathcal{E}^\prime_n$. The algebraic 
element $: \varphi^{\otimes n}(t) :_\omega$ 
is just the smeared $n$-th normal ordered Wick power of 
the free field at the same 
spacetime point, 
\begin{equation}
\label{wickdef}
:\varphi^n(f):_\omega \,\,= \,\,
: \varphi^{\otimes n}(t) :_\omega
, 
\end{equation}
as previously defined in \cite{bfk}. More generally, it can be 
shown \cite{bf}, that $\mathcal W$ also contains time-ordered 
products of normal ordered Wick-products. 

Using Wick's theorem, one can show that the product of 
two elements in $\mathcal W$ of the form \eqref{Adef} can 
again be written as a finite sum of elements of this form. 
This shows in particular that any element in $\mathcal W$ 
arises as a finite sum of elements of the form \eqref{Adef}
with $t_n \in \mathcal{E}_n^\prime$ plus a multiple of the 
identity operator.

\medskip

For later purposes, we would like to have a suitable notion 
of the continuity of states on the algebras $\mathcal W$ and $\mathcal A$. 
In order to define such a notion, we must first equip $\mathcal W$
(and therefore also $\mathcal A$) with a topology. In other words, we must 
explain what we mean by the statement that a sequence $\{A_\kappa\}$ of 
elements in $\mathcal W$ converges to an element $A$. Such a 
topology has been defined in \cite{hw}, we here briefly indicate
the main idea. One first defines a notion of convergence of 
a sequence $\{t_\kappa\}$ in the spaces of distributions 
$\mathcal{E}^\prime_n$ defined above in \eqref{edef}. 
Namely, such a sequence is said to converge to a distribution $t$ if 
\begin{enumerate}
\item[(a)] 
the support of $t_\kappa$ is contained in some compact set 
$K$ for all $\kappa$, 
\item[(b)]
$t_\kappa \to t$ weakly in the sense of distributions, 
\item[(c)] there 
is a closed conic set $\Gamma \subset T^*M^n \backslash (V_+^n 
\cup V_-^n)$ such that ${\rm WF}(t_\kappa) \subset \Gamma$ for all 
$\kappa$, 
\item[(d)] for any properly supported pseudo differential 
operator $P$ with $\mu{\rm supp}(P) \cap \Gamma = \emptyset$, we have that 
$Pt_\kappa \to Pt$ in the sense of compactly supported smooth functions. 
\end{enumerate}

\paragraph{Remark.}
It is common to say that a sequence of distribtuions $\{t_\kappa\}$ satisfying 
(b) through (d) ``converges to $t$ in 
the sense of $\cD'_\Gamma$''. For an explanation of the 
notion of a pseudo differential operator and 
the related technical terms appearing in item (d), we refer the 
reader to \cite{h}. It can be shown that $t$ is again an 
element in $\mathcal{E}'_n$, so these spaces
are complete with respect to the above topology.

\medskip

Having defined a notion of sequential
convergence within the spaces $\mathcal{E}'_n$, 
we now define a notion of sequential convergence in the algebra 
$\mathcal W$ as follows. Let  $\{A_\kappa\}$ be a sequence 
of generators in $\mathcal W$, defined by distributions 
$t_\kappa \in \mathcal{E}'_n$ as in \eqref{Adef}. Then we say 
that the sequence $\{A_\kappa\}$ converges to an element $A$ of the 
form \eqref{Adef}, if $t_\kappa \to t$ in $\mathcal{E}'_n$.
The so defined notion of covergence for the generators of 
$\mathcal W$ generalizes to arbitrary sequences in $\mathcal{W}$,
because every element of this algebra can be written as a 
finite linear combination of the generators.

A state $\psi$ on $\mathcal{W}$ is said to 
be continuous, if $\psi(A_\kappa) \to \psi(A)$ whenever
$A_\kappa$ converges to $A$. We note that, since the 
space of smooth testfunctions on $M^n$ is dense in the 
space $\mathcal{E}^\prime_n$, the algebra $\mathcal A$ 
is dense in $\mathcal W$ in the above topology. Therefore 
we have the important result that continuous states 
on $\mathcal W$ are completely determined by their restrictions
to $\mathcal A$. If we consider sequences $\{t_\kappa\}$ 
satifying (a) through (d) with $\Gamma = \emptyset$, then
we get that the $n$-point functions of a continuous 
state must be continuous in the Laurent-Schwartz topology
on the space of smooth testfunctions. Thus, we find in 
particular that the $n$-point functions of a 
continuous state are distributions.

\section{The state space of $\mathcal W$}
\label{sec3}
The aim of this section is to characterize the space 
of continuous states on the algebra $\mathcal W$. We first note that, 
since $\mathcal{W} \supset \mathcal{A}$, every 
continuous state $\psi$
on $\mathcal{W}$ gives rise, by restriction, to a state 
on the minimal algebra $\mathcal A$ whose $n$-point functions are 
distributions. However, the opposite is 
not true, namely it is not true that every such state on $\mathcal A$
can be extended to a continuous state on $\mathcal W$. 
This may be seen for example by considering 
the smeared normal ordered Wick-power 
$: \varphi^2 (f) :_\omega$, which is an 
element of $\mathcal{W}$, but which is not an element 
of $\mathcal A$. Now if $\psi$ is a state on $\mathcal A$
with distributional $n$-point functions, then its 
action---provided it can be defined---on this Wick-power must 
be given by the limit
\begin{equation}
\label{example}
\psi(: \varphi^2(f) :_\omega) = \lim_{\kappa \to \infty} \int 
(\psi_2 - \omega_2)(x_1, x_2) f(x_1) \delta_\kappa (x_1, x_2) \,d\eta_1
d\eta_2, 
\end{equation}
where $\{\delta_\kappa\}$ is a suitable sequence of smooth functions 
tending to the delta-distribution. (Note that this prescription 
is just a reformulation of the usual ``point-splitting'' method, 
as explained for example in \cite{wald}.)
However, this limit will only exist and be independent of the 
particular choice of sequence $\{\delta_\kappa\}$
if the distribution $(\psi_2 - \omega_2)(x_1, x_2)$ is  
at least continuous at $x_1 = x_2$. 
There are many states on $\mathcal A$
which do not have this property and which therefore do not 
extend to $\mathcal W$.

\medskip

The precise characterization of the space of continuous states
on $\mathcal W$ is as follows:

\begin{thm}
\label{mainthm}

\begin{enumerate}
\item[(i)]
Let $\psi$ be a continuous state  on $\mathcal{W}$. 
Then the two-point function of the free field 
must be of Hadamard form and the truncated $n$-point functions
of the free field must be smooth for $n \neq 2$. 

\item[(ii)]
Conversely, if $\psi$ is a state on $\mathcal A$ whose 
two-point function is of Hadamard form and whose truncated
$n$-point functions are smooth for all $n \neq 2$, then 
$\psi$ extends to a (necessarily unique) continuous 
state on $\mathcal W$.
\end{enumerate}
\end{thm}

\paragraph{Remarks.}
\begin{enumerate}

\item
The results of 
\cite{bf, hw} imply that any quasi-free Hadamard state on 
$\mathcal A$ can be extended to a 
continuous state on $\mathcal W$. Clearly, this is a 
special case of item (ii) of the above theorem, since 
quasi-free states by definition have vanishing truncated 
$n$-point functions for $n \neq 2$.

\item
B. S. Kay has shown (unpublished manuscript) that the $N$-particle 
states with smooth mode functions in the GNS-representation of any 
quasi-free Hadamard state are Hadamard and have smooth truncated 
$n$-point functions for $n \neq 2$. By (ii) of our theorem, 
these states therefore extend to continuous states on $\mathcal W$.

\item
On Robertson-Walker spacetimes, there exists the notion of 
``adiabatic vacuum states'' on $\mathcal A$, introduced by 
Parker and defined in a mathematically rigorous
way by \cite{lr}. The two-point function of such a  state differs
from that of a Hadamard state typically by a term which is a number 
of times differentiable, but which is not smooth \cite{ho}. 
By our theorem, adiabatic states therefore do not possess an extension
to continuous states on $\mathcal W$. The same remark applies to the 
class of states recently introduced by Junker \cite{j}. 

\item
It is known \cite{v} that all quasifree Hadamard states on 
$\mathcal A$ are ``locally
quasi-equivalent'', that is, any quasifree Hadamard state locally
arises as a density matrix state in the GNS-representation of 
any other such state.\footnote{Actually, in \cite{v}, the author 
does not work with the minimal algebra $\mathcal A$, but instead
with the algebra generated by the exponentiated smeared 
fields. However, 
this difference is not relevant in as far as the issue of 
quasi-equivalence is concerned.}
We here conjecture that, more generally, 
this is true for all states described in item (ii) above. 
In view of the above theorem, this would imply that any two
continuous states on $\mathcal W$ are locally quasi-equivalent.

\item
In \cite{bfk}, the authors introduce a ``microlocal spectrum condition'', 
which generalizes to curved spacetimes 
the usual spectrum condition imposed on the $n$-point
function of a field theory in the context of the Wightman-axioms. 
The content of this condition is to require that 
wave front set of the $n$-point functions of an 
admissible  state should have a specific form. 
The microlocal spectrum condition is 
known to hold for the $n$-point functions of the free field
in any quasi-free Hadamard state. It is an easy consequence
of our result that the microlocal spectrum condition holds 
in fact for the $n$-point functions of any continuous state on 
$\mathcal W$.  

\item
We shall from now on only deal with continuous states. Therefore, for 
simplicity, whenever we speak of ``states'', we shall 
mean ``continuous states''. 

\end{enumerate}

\paragraph{Proof.}

We begin with the proof of (i). 
Let thus $\psi$ be a continuous state on $\mathcal W$. 
We need to show that the two-point function, $\psi_2$,
is of Hadamard form, and that the truncated $n$-point functions, 
$\psi_n^T$, are all smooth, except for $n=2$. Let
\begin{eqnarray}
\label{psidef}
\Psi_n( f_1, f_2, \dots , f_n) &\equiv &
\psi \left( :\varphi(f_1) \varphi(f_2)\dots \varphi(f_n) 
:_{\omega }\right). 
\end{eqnarray}
Then we have 
\begin{lem}
\label{start}
Let $\psi$ be a state on $\mathcal A$. Then 
the following statements are equivalent: 
\begin{enumerate}
\item[(i)]
$\psi_2$ is Hadamard and $\psi_n^T$ are smooth for $n \neq 2$.
\item[(ii)]
The distributions $\Psi_n$ are smooth for all $n$. 
\end{enumerate}
\end{lem}
\paragraph{Proof.} 
The proof is based on the following combinatorical
formula, which we shall prove in the appendix:
\begin{equation}
\label{eq2}
(\psi^T_n - \omega^T_n)(f_1, f_2, \dots, f_n)
= 
\frac{\partial^n}{i^n \partial t_1 \partial t_2 \dots \partial t_n}
\ln \psi\left( G(\sum_i t_i f_i) \right)  \Bigg|_{t_1 = 
t_2 = \dots = 0}, 
\end{equation}
where $G(f)$ is defined in \eqref{gdef}. If one carries out the  
differentiations in formula \eqref{eq2} and uses that 
the functional $f \to \psi(G(f))$ is the generating functional for the 
hierarchy of distributions $\{\Psi_1, \Psi_2, \dots, \Psi_n, \dots \}$,
as well as the standard relation $\ln(1 + x) = 
\sum_{k\ge 1} (-1)^{k+1} x^k/k$, then one obtains the formula
\begin{equation}
\label{eq3}
(\psi^T_n - \omega^T_n)(f_1, f_2, \dots, f_n) = 
\sum_{P \in \mathcal{I}_n} 
(-1)^{|P|-1}(|P|-1)! 
\prod_{r \in P} \Psi_{|r|}(f_{r(1)}, f_{r(2)}, \dots, f_{r(|r|)}).
\end{equation}
Thus, if $\Psi_n$ is smooth for all $n$, 
then so is $\psi_n^T - \omega_n^T$. For $n=1$
this means that $\psi_1$ is smooth. For $n=2$ this means that
$\psi_2 - \omega_2 - \psi_1 \otimes \psi_1$ is smooth, and hence that
$\psi_2$ is Hadamard. For $n\ge 3$ this shows that $\psi_n^T$ is
smooth, since $\omega_n^T = 0$ for all $n \ge 3$. We have thus 
shown the implication $(i) \Longrightarrow (ii)$ of the Lemma. 
The implication $(ii) \Longrightarrow (i)$ can be shown similarly
by solving \eqref{eq3} for $\Psi_n$ in terms of $\psi^T_k -
\omega_k^T$ with $k \le n$. \qed

\medskip

It thus remains to be 
shown that $\Psi_n$ is smooth for all $n$. 
We begin by showing that the wave front set  
of $\Psi_n$ is not arbitrary.

\begin{lem}
\label{lem0}
Let $\psi$ be a continuous state on $\mathcal{W}$. Then
necessarily
\begin{equation}
{\rm WF}\left( \Psi _{n}\right) \subset V_+^n \cup V_-^n
\quad \text{for all $n$.}
\label{intro1}
\end{equation}
\end{lem}
\paragraph{Proof.}
Given in the Appendix. \qed

In order to show that the wave front set of the distributions
$\Psi_n$ is in fact empty, we proceed by an induction in $n$.
For $n=0$ there is nothing to prove, 
since $\Psi_0 = 1$, which is clearly smooth. Let us therefore assume 
that $\Psi_k$ is smooth for all $k \le n-1$. We need to prove that also 
$\Psi_n$ is smooth. For this, it is necessary to gain some 
information about the Fourier transform, 
$\widehat {\chi_n \Psi_n}(l_1, \dots, l_n)$, 
in directions such that either all $l_i$ are in the future 
lightcone at some points $x_i$ or all $l_i$ are in the past light cone 
of some points $x_i$, where $\chi_n$ is a smooth 
bump function whose support is 
localized around the point $(x_1, \dots, x_n)$ in the product 
manifold $M^n$. We prepare the ground  with the following 
three lemmas.

\begin{lem}
\label{lem3}
Let $\psi $ be a state on $\mathcal{A}$. 
Then 
there holds 
\begin{equation}
\left| \psi _{n}\left( f_{1},\ldots ,f_{n}\right) \right| ^{2}\leq 
\psi_{2n}\left( f_{1},\ldots ,f_{n},\bar{f}_{n},\ldots ,\bar{f}_{1}\right), 
\end{equation}
for all $n$ and all testfunctions. 
\end{lem}

\paragraph{Proof.}

By the the Cauchy-Schwartz inequality
\begin{equation}
\left| \psi( A ) \right| ^{2}\leq \psi(AA^*)
\end{equation}
for all $A \in \mathcal{A}$. The statement
of the lemma then follows by setting  $A=\varphi \left( f_{1}\right)
\cdots \varphi \left( f_{n}\right) $. \qed

For the next lemma, we introduce 
the following notation. We denote by $P$ a partition of the 
set $\{1, \dots, n\}$ into disjoint ordered pairs 
$\{(i_1, j_1), \dots (i_{|P|}, j_{|P|}) \}$, meaning 
that $i < j$ for all $(i, j) \in P$. 
The number of pairs in the partition $P$ is denoted by $|P|$. 
The set of all such partitions for a given $n$ is 
denoted by ${\cal P}_n$. If $1 \le k \le n$, 
then we write $k \in P$ if the partition $P$ contains a pair $(i, j)$
such that either $k=i$ or $k=j$.  

\begin{lem}
\label{lem1}
Let $\psi $ be a state on $\mathcal{A}$. 
Then there holds 
\begin{equation}
\label{psn}
\psi_n(f_1, f_2, \dots, f_n) 
= \sum_{P \in {\cal P}_n} (-1)^{|P|}
\Psi_{n - 2|P|}(\otimes_{k \notin P} f_k)
\prod_{(i,j) \in P} \omega_2(f_i, f_j)
\end{equation}
for all testfunctions $f_1, f_2, \dots , f_n $. 
\end{lem}
\paragraph{Proof.}
Recall that $G(f)$, defined in \eqref{gdef},
is the generating functional for the Wick products 
$: \varphi(x_1) \varphi(x_2) \dots \varphi(x_n) :_\omega$. Therefore
$\psi(G(f))$ is the generating functional for the distributions 
$\Psi_n$. By repeatedly using the 
identity 
\begin{equation}
e^{i\varphi(f_1)} e^{i\varphi(f_2)} = 
e^{i\varphi(f_1 + f_2)} 
e^{-\frac{i}{2}\Delta(f_1, f_2)},  
\end{equation}
it is straightforward
to calculate that 
\begin{equation}
\psi\left(e^{it_1\varphi(f_1)} \dots e^{it_n\varphi(f_n)} \right)=
\exp\left(\sum_{i<j} t_i t_j \omega_2(f_i, f_j) 
+ \frac{1}{2} \sum_i t_i^2 \omega_2(f_i, f_i) \right)
\psi\left( G(\sum_i t_i f_i) \right).  
\end{equation}
Applying $(-i)^n \partial^n/\partial t_1 \dots \partial t_n$ to 
both sides of this equation and setting $t_1, \dots, t_n$ to 
zero then yields the formula claimed in the lemma. 
\qed
\begin{lem}
\label{lem4}
Let $\psi$ be a continuous state on $\mathcal{W}$, and 
let $n \ge 1$. Then ${\rm WF}(\psi_{2n})$ does not contain any 
elements of the form 
\begin{equation}
(x_1, k_1, \dots, x_n, k_n, x_n, -k_n, \dots, x_1, -k_1)
\quad \text{with $k_i \in (V_-)_{x_i}$ for all $i$.}
\end{equation}
\end{lem}
\paragraph{Proof.} 
As a preparation, let us start by introducing some notation. 
Let $(i_1, \dots, i_r)$ be tuple of natural numbers with 
$1 \le i_1 < i_2 < \dots i_r \le 2n$. For each such a tuple, we 
define a map $\phi_{(i_1, i_2, \dots, i_r)} : M^{2n} \to M^r$ by 
\begin{equation}
\phi_{(i_1, i_2, \dots, i_r)}(x_1, x_2, \dots, x_{2n}) \equiv
(x_{i_1}, x_{i_2}, \dots, x_{i_r}).
\end{equation}
With this notation, Eq. \eqref{psn} can be rewritten as 
\begin{equation}
\label{psn1}
\psi_{2n}
= \sum_{P \in {\cal P}_{2n}} (-1)^{|P|} 
\phi^*_{(k_1, \dots, k_{2n-2|P|})} 
\Psi_{n - 2|P|} \cdot
\prod_{(i,j) \in P} \phi^*_{(i,j)} \omega_2,
\end{equation}
where $\{k_1, \dots, k_{2n-2|P|}\}$ is the set 
of numbers in $\{1, \dots, 2n\}$ which are not contained 
in the partition $P$, and where the pull-back of a 
distribution is defined by analogy with the pull-back of 
a smooth density. Note that the distributions $\phi_{(i,j)}^*
\omega_2$ etc. are by definition distributions on $M^{2n}$, 
so the products in formula \eqref{psn1} denote the pointwise
product of distributions on $M^{2n}$.

Using now formulas \cite[I, Thms. 8.2.10 and 8.2.4]{h} for 
the wave front of products and pull-backs of distributions, we 
get the estimate
\begin{eqnarray}
{\rm WF}(\psi_{2n})
&\subset& \bigcup_{P \in \mathcal{P}_{2n}}
{\rm WF}\left(\phi^*_{(k_1, \dots, k_{2n-2|P|})} 
\Psi_{n - 2|P|} \cdot
\prod_{(i,j) \in P} \phi^*_{(i,j)} \omega_2 \right)\nonumber\\
&\subset& \bigcup_{P \in \mathcal{P}_{2n}}
\left[ \Big\{\phi^*_{(k_1, \dots, k_{2n-2|P|})} 
{\rm WF}\left(
\Psi_{n - 2|P|} \right) \cup \{0\} \Big\} + 
\sum_{(i,j) \in P} \Big\{ \phi^*_{(i,j)} {\rm WF}\left(
\omega_2 
\right) \cup \{0\}\Big\}
\right]\nonumber\\
&\subset&
\bigcup_{P \in \mathcal{P}_{2n}}
\left[
\phi^*_{(k_1, \dots, k_{2n-2|P|})} (V_+^{2n-2|P|} \cup
V_-^{2n-2|P|})+ 
\sum_{(i,j) \in P} \phi^*_{(i,j)} (V_+ \times V_-)
\right], 
\end{eqnarray} 
where we have used that ${\rm WF}(\omega_2) \subset V_+ \times V_-$
since $\omega$ is Hadamard, and that ${\rm WF}(\Psi_k) \subset
V_+^k \cup V_-^k$, by Lem. \ref{lem0}. Hence, in order to prove the lemma, 
it is sufficient to demonstrate that if a vector $(\underline x, \underline
l) \in T^*M^{2n}$
of the form 
\begin{equation}
\label{lform}
(x_1, l_1, \dots, x_{2n}, l_{2n}) \equiv
(x_1, k_1, \dots, x_n, k_n, x_n, -k_n, \dots, x_1, -k_1) 
\quad \text{with $k_i \in (V_-)_{x_i}$ for all $i$} 
\end{equation}
is in the set
\begin{equation}
\label{set}
\phi^*_{(k_1, \dots, k_{2n-2|P|})} (V_+^{2n-2|P|} \cup
V_-^{2n-2|P|})+ 
\sum_{(i,j) \in P} \phi^*_{(i,j)} (V_+ \times V_-)
\end{equation}
for some partition $P$, then $k_i = 0$ for all $i$. 
So let $(\underline x, \underline l)$ be in 
the set \eqref{set} for some $P$. This implies that
\begin{enumerate}
\item[(a)]
$l_i \in V_-, l_j \in V_+$ for all $(i,j) \in P$. 
\item[(b)]
Either $l_i \in V_+$ for all $i \notin P$ or 
$l_i \in V_-$ for all $i \notin P$. 
\end{enumerate}
Now property (a), together with the specific form of 
$(\underline x, \underline l)$ given by \eqref{lform}, 
implies that $l_i = l_j = 0$
whenever $(i, j) \in P$. Combining this with property (b), 
we see that the $l_i$ must either be all in $V_-$ or all 
in $V_+$. Using again the specific form of $(\underline x, \underline l)$, 
we conclude that this is only possible when all $l_i = 0$, 
implying that all $k_i = 0$.
\qed

\medskip

We have now gathered enough information to show that $\Psi_n$ is 
smooth for all $n$. By the induction hypothesis, we 
know that $\Psi_k$ is smooth for all $k \le n-1$. 
We want to use this to 
obtain an estimate for the Fourier transform of $\chi_n\Psi_n$, 
where $\chi_n$ is a smooth function with compact support
that will be specified momentarily.

From Lemmas \ref{lem1} and \ref{lem3} we get the inequality
\begin{multline}
\psi_{2n}(f_1, \dots, f_n, \bar f_n, \dots, \bar f_1) \ge\\
\sum_{P, P' \in {\cal P}_{n}}
(-1)^{|P|+|P'|} \Psi_{n-2|P|}(\otimes_{k \notin P} \bar f_k)
\Psi_{n-2|P'|}(\otimes_{k' \notin P'} f_{k'})
\prod_{(i,j) \in P} \omega_2(\bar f_j, \bar f_i) 
\prod_{(i',j') \in P'} \omega_2(f_{i'}, f_{j'}).
\end{multline}
We now specialize the above inequality to 
testfunctions $f_j$ of the form 
\begin{equation}
f_j(x) = \frac{1}{(2\pi)^2} \eta_j(x) e^{il_j x}. 
\end{equation}
Here, $\eta_j$ are real-valued smooth bump functions whose support is 
contained in some chart, $l_j$ are vectors in $\mr^4$ and the expression 
$l_j x$ denotes the scalar product in $\mr^4$ between $l_j$ 
and the coordinate
components of $x$ in the above chart. With this 
choice for $f_j$, the above inequality can be rewritten as 
\begin{multline}
\label{master}
|\widehat{\chi_n\Psi_n}({l_1}, \dots, {l_n})|^2 \le
\widehat{\chi_{2n} 
\psi_{2n}}({l_1}, \dots, {l_n}, {-l_n}, \dots, {-l_1}) -
\\
\sum_{P, P' \in {\cal P}_{n}, P, P' \neq \emptyset}
(-1)^{|P|+|P'|} \widehat {\chi_{n-2|P|} \Psi_{n-2|P|}}\left(-l_{k_1}, 
\dots, -l_{k_{n-2|P|}}\right)
\widehat 
{\chi_{n-2|P'|} \Psi_{n-2|P'|}}\left(l_{k_1'}, 
\dots, l_{k_{n-2|P'|}'}\right)\\
\times \prod_{(i,j) \in P} \widehat {\chi_2 \omega_2}({-l_j}, {-l_i}) 
\prod_{(i',j') \in P'} \widehat {\chi_2\omega_2}({l_{i'}}, 
{l_{j'}}),
\end{multline}
where $\{k_1, \dots, k_{2n-2|P|}\}$ is the set 
of numbers in $\{1, \dots, n\}$ which are not contained 
in the partition $P$ and where $\{k_1', \dots, k_{2n-2|P'|}'\}$ is the set 
of numbers in $\{1, \dots, n\}$ which are not contained 
in the partition $P'$. The smooth functions $\chi_k$ denote suitable 
tensor products of $k$ factors of the functions $\eta_i$. For example, in 
the expression $\widehat{\chi_2 \omega_2}(l_{i'}, l_{j'})$, the 
function $\chi_2$ should be taken to be $\chi_2 = \eta_{i'} \otimes
\eta_{j'}$; in the expression $\widehat{\chi_{2n} 
\psi_{2n}}({l_1}, \dots, {l_n}, {-l_n}, \dots, {-l_1})$, the function 
$\chi_{2n}$ should be taken to be $\chi_{2n} = \eta_1 \otimes \dots
\eta_n \otimes \eta_n \otimes \dots \eta_1$, etc.

We would now like to argue that the right side of inequality 
\eqref{master} is rapidly decaying in directions for which  
all $l_i$ are in some conic 
neighborhood of $(V_-)_{x_i}$, with at least one $l_i
\neq 0$, provided the support of the functions $\eta_i$ is 
localized sharply enough around points $x_i$. For this, we first look 
at the terms in the sum on the right side of inequality \eqref{master}. 
Each term in this sum contains at least one factor of either the form 
$\widehat {\chi_2\omega_2}(l_i, l_j)$ or 
$\widehat {\chi_2\omega_2}(-l_{j'}, -l_{i'})$, 
where $(i, j) \in P$ or 
$(i', j') \in P'$. Provided that not all the covectors 
$l_i, l_j, l_{i'}, l_{j'}$ occurring in these factors are zero, 
these factors give us rapid decay of the corresponding term in 
the sum. This is because, by the Hadamard property of $\omega_2$,  
\begin{equation}
|\widehat{\chi_2 \omega_2}(k_1, k_2)| \le
C_N(1 + |k_1| + |k_2|)^{-N} 
\end{equation}
for all $N$ and all directions $(k_1, k_2)$  in some 
conic neighborhood of $(V_+)_{x_1} \times
(V_+)_{x_2} \cup (V_-)_{x_1} \times (V_-)_{x_2}$, provided 
$\chi_2$ is localized sufficiently sharply around $(x_1, x_2)$.  
If all the $l_i, l_j, l_{i'}, l_{j'}$ occurring in the factors
$\widehat {\chi_2\omega_2}(l_i, l_j)$ or 
$\widehat {\chi_2\omega_2}(-l_{j'}, -l_{i'})$ are zero, then 
at least one of the covectors $l_k$ with $k \notin P$ 
and at least one of the covectors $l_{k'}$ with 
$k' \notin P'$ must be nonzero, since otherwise all the 
$l_i$ would be zero. Let us first assume that $P$ is not 
the empty set. Then we have
\begin{equation}
\Big|\widehat {\chi_{n-2|P|} \Psi_{n-2|P|}}\left(-l_{k_1}, -l_{k_2},
\dots, -l_{k_{n-2|P|}}\right)\Big| \le 
C_N\left(1+\sum_{k \notin P} |l_k|\right)^{-N}
\end{equation} 
for all $N$ and suitable constants $C_N$, since 
the distributions $\Psi_{n-2|P|}$ are smooth by the inductive 
assumption. If $P = \emptyset$, 
then $P'$ is not the empty set, and we get an estimate of the 
above form for the term $\widehat 
{\chi_{n-2|P'|} \Psi_{n-2|P'|}}$. In summary, we have shown that 
each term in the sum on the right side of \eqref{master} contains
at least one factor which is rapidly decaying. Therefore the 
whole sum is rapidly decaying in directions such that either all 
$l_i \in (V_+)_{x_i}$ or all $l_i \in (V_-)_{x_i}$  
and not all $l_i = 0$, provided the functions $\eta_i$ are localized 
sufficiently sharply around the points $x_i$. 

By Lemma \ref{lem4}, 
the first term on the right side of \eqref{master} is rapidly decaying 
in directions for which all $l_i$ are in a conic 
neighborhood of $(V_-)_{x_i}$, 
provided the functions 
$\eta_i$ are localized sharply enough around $x_i$. 
Hence, we have altogether
found that 
\begin{equation}
\label{minq}
|\widehat{\chi_n\Psi_n}({l_1}, l_2, \dots, {l_n})| \le
C_N\left(1 + \sum_i |l_i|\right)^{-N}
\end{equation}
in directions for which all $l_i$ are in a neighborhood 
of $(V_-)_{x_i}$, provided the functions 
$\eta_i$ are localized sharply enough around $x_i$. Now, since 
\begin{equation}
(:\varphi^{\otimes n} (t) :_\omega)^* \,\, 
= \,\,:\varphi^{\otimes n} (\bar t) :_\omega,
\end{equation}
for all testfunctions $t$, 
the distributions $\Psi_n$ are real, in the sense that 
$\overline{\Psi_n(t)} = \Psi_n(\bar t)$ for all 
testfunctions $t$. Hence, $|\widehat{\chi_n\Psi_n}(l_1, 
\dots, l_n)| = |\widehat{\chi_n\Psi_n}(-l_1, 
\dots, -l_n)|$, and the inequality \eqref{minq} must therefore
also hold if all the $l_i$ in that inequality 
are replaced by $-l_i$, that is, \eqref{minq} must also hold 
in directions for which all $l_i$ are in some neighborhood
of the cones $(V_+)_{x_i}$. Therefore, since 
we already know that ${\rm WF}(\Psi_n) \subset V_+^n \cup V_-^n$, 
we get from this that the distribution $\chi_n \Psi_n$ has in fact no 
singular directions at all, provided the supports of $\eta_i$ are 
sufficiently sharply localized around $x_i$. But this implies that
${\rm WF}(\Psi_n) = \emptyset$, as we wanted to show. 

\vspace{1cm}

We next prove (ii). Let us 
thus assume $\psi$ is a state on $\mathcal A$ 
for which $\psi_2$ is Hadamard and for which $\psi_n^T$ are
smooth for all $n \neq 2$. By Lemma \ref{start}, this 
implies that the distributions $\Psi_n$ are smooth. We are thus 
allowed to define an action of $\psi$ on elements of
$\mathcal{W}$ by the formula
\begin{equation}
\psi(: \varphi^{\otimes n}(t) :_\omega) \equiv \Psi_n(t),
\end{equation}
for all $t \in \mathcal{E}'_n$ and all $n$. It is easily checked that
this formula defines a linear, normalized and 
continuous  functional on $\mathcal{W}$
which extends the action of $\psi$ on $\mathcal{A}$. Moreover, 
this functional is also positive, since it is continuous 
and positive on $\mathcal A$, which is a dense subspace
of $\mathcal W$.
\qed

\section{Appendix} 

\subsection{Proof of formula \eqref{eq2}}

Let $\{h_1, h_2, \dots, h_n, \dots \}$ denote some hierarchy of 
symmetric distributions and let 
\begin{equation}
H(f) = 1 + \sum_{n \ge 1} \frac{i^n}{n!} h_n(f, f, \dots, f) 
\end{equation}
be the corresponding generating functional. Then the 
``linked cluster theorem'' (see e.g. \cite[pp 125]{hdp})
states that the corresponding 
truncated distributions are given by 
\begin{equation}
h^T_n(f_1, f_2, \dots, f_n) 
= 
\frac{\partial^n}{i^n \partial t_1 \partial t_2 \dots \partial t_n}
\ln H (\sum_i t_i f_i) \Bigg|_{t_1 = 
t_2 = \dots = 0}. 
\end{equation}
We would like to apply this result to the hierarchies 
$\{\psi_1, \psi_2, \dots, \psi_n, \dots \}$ and
$\{\omega_1, \omega_2, \dots, \omega_n, \dots \}$ of the 
$n$-point functions of the states $\psi$ and $\omega$.
However, these are not symmetric and therefore the 
linked cluster theorem is not directly applicable. 
Instead, we first apply the linked cluster theorem 
to the hierarchies of symmetrized $n$-point functions, 
$\{\psi_1^S, \psi_2^S, \dots, \psi_n^S, \dots \}$ and
$\{\omega_1^S, \omega_2^S, \dots, \omega_n^S, \dots \}$, 
where the superscript ``$S$'' stands for symmetrization.
This gives us  
\begin{eqnarray}
\label{eq4}
[(\psi^S_n)^T - (\omega^S_n)^T](f_1, \dots, f_n)
&=& 
\frac{\partial^n}{i^n \partial t_1 \dots \partial t_n}
\left[
\ln \psi\left(e^{i\varphi(\sum_i t_i f_i)}\right)  
- \ln \omega\left(e^{i\varphi(\sum_i t_i f_i)}\right)  
\right]
\Bigg|_{t_1 = 
t_2 = \dots = 0}\nonumber\\
&=&\frac{\partial^n}{i^n \partial t_1 \dots \partial t_n}
\ln \psi\left(G(\sum_i t_i f_i) \right)
\Bigg|_{t_1 = 
t_2 = \dots = 0}, 
\end{eqnarray}
where we have used the definition of $G(f)$, Eq. \eqref{gdef}, 
as well as the relation $\omega(e^{i\varphi(f)}) = e^{-\frac{1}{2}
\omega_2(f, f)}$, which holds because 
$\omega$ is quasi-free. The desired relation \eqref{eq2} then
follows if we can show that
\begin{equation}
\label{*}
(\psi^S_n)^T - (\omega^S_n)^T = \psi_n^T - \omega_n^T
\end{equation}
for all $n$. The demonstration of \eqref{*} makes up 
the rest of this subsection. 

Relation \eqref{*} can be checked immediately for $n = 1$ and 
$n = 2$. For $n \ge 3$ it reduces to 
\begin{equation}
\label{**}
(\psi^S_n)^T = \psi_n^T 
\end{equation}
since $\omega^T_n = (\omega^S_n)^T = 0$ for $n \ge 3$. 
In order to see \eqref{**}, we first note that the 
truncated $n$-point functions of any state on $\mathcal{A}$
are symmetric for $n \ge 3$, $\psi_n^T = (\psi_n^T)^S$, 
as one can show by a straightforward inductive argument
using the commutation relation \eqref{com} for the free field. 
Eq. \eqref{**} thus follows from the fact that for any
hierarchy $\{h_1, h_2, \dots, h_n, \dots \}$ (not necessarily
symmetric) there holds
\begin{equation}
\label{eq5}
(h_n^T)^S = (h_n^S)^T
\end{equation}
for all $n$. To see this, we argue as follows. 
Let $P=\left\{ r_{1},\ldots ,r_{k}\right\} $, and let $Q\left( r_{j}\right) $
denote the set of all permutations of $r_{j}\equiv \left( r_{j}\left(
1\right) ,\ldots ,r_{j}\left( n_{j}\right) \right) $ where $n_{j}=\left|
r_{j}\right| $. Let
\[
Q\left( P\right) =\left\{ \sigma _{1}\cdots \sigma _{k}|\hspace{5pt}\sigma
_{j}\in Q\left( r_{j}\right) \hspace{5pt}\text{for }j=1,\ldots ,k\right\} .
\]
Let $\mathcal{I}\left(
n_{1},\ldots ,n_{k}\right) $ denote the subset of partitions which has $k$
members $\left\{ r_{1},\ldots ,r_{k}\right\} $ such that for each $%
i=1,\ldots ,k$, $\left| r_{j}\right| =n_{j}$ for all $j$. For any fixed $P\in 
\mathcal{I}\left( n_{1},\ldots ,n_{k}\right) $, we then get 
\begin{eqnarray*}
\left( \prod_{r\in P}h_{\left| r\right| }\right) ^{S}\left( x_{1},\ldots
,x_{n}\right)  &=&\frac{1}{n!}\sum_{P^{\prime }\in \mathcal{I}\left(
n_{1},\ldots ,n_{k}\right) }\sum_{\sigma \in Q\left( P^{\prime }\right)
}\left( \prod_{r_{j}\in P^{\prime }}h_{n_{j}}\right) \left( x_{\sigma \left(
1\right) },\ldots ,x_{\sigma \left( n\right) }\right)  \\
&=&\frac{n_{1}!\cdots n_{k}!}{n!}\sum_{P^{\prime }\in \mathcal{I}\left(
n_{1},\ldots ,n_{k}\right) }\prod_{r_{j}\in P^{\prime }}h_{n_{j}}^{S}\left(
x_{r_{j}\left( 1\right) },\ldots ,x_{r_{j}\left( n_{j}\right) }\right) .
\end{eqnarray*}
(Note that the left hand side is independent of $P\in \mathcal{I}%
\left( n_{1},\ldots ,n_{k}\right) $.) From this we conclude that
\begin{eqnarray}
\label{cmp}
\left( \sum_{P\in \mathcal{I}_{n}}\prod_{r\in P}h_{\left| r\right| }\right)
^{S}\left( x_{1},\ldots ,x_{n}\right)  &=&\sum_{\left\{ n_{1},\ldots
,n_{k}\right\} }\sum_{P\in \mathcal{I}\left( n_{1},\ldots ,n_{k}\right)
}\left( \prod_{r\in P}h_{\left| r\right| }\right) ^{S}\left( x_{1},\ldots
,x_{n}\right)  \nonumber\\
&=&\sum_{\left\{ n_{1},\ldots ,n_{k}\right\} }\sum_{P^{\prime }\in 
\mathcal{I}\left( n_{1},\ldots ,n_{k}\right) }\prod_{r_{j}\in P^{\prime
}}h_{n_{j}}^{S}\left( x_{r_{j}\left( 1\right) },\ldots ,x_{r_{j}\left(
n_{j}\right) }\right)  \nonumber\\
&=&\sum_{P\in \mathcal{I}_{n}}\prod_{r\in P}h_{n_{j}}^{S}\left(
x_{r_{j}\left( 1\right) },\ldots ,x_{r_{j}\left( n_{j}\right) }\right) 
\end{eqnarray}
where the sum $\sum_{\left\{ n_{1},\ldots ,n_{k}\right\} }$ is over all
possible set of positive integers such that $n_{1}+\ldots +n_{k}=n$.

\medskip

We now use this formula to prove Eq. \eqref{eq5}. 
Suppose that this equation is true for $1, 2, \dots, n$. 
We now show that it must also be true for $n+1$. 

By the induction hypothesis, for any $P\in \mathcal{I}%
_{n+1}$ which is not $P_0=\{(1, 2, \dots, n+1)\}$ and any 
$r\in P$, we have that 
\begin{equation}
\label{induct}
(h_{\left| r\right| }^S)^T
\left( x_{r\left( 1\right) },\ldots ,x_{r\left(
\left| r\right| \right) }\right) = ( h_{\left| r\right| }^T)^{S}
\left( x_{r\left( 1\right) },\ldots ,x_{r\left( \left| r\right| \right)
}\right) .
\end{equation}
Hence, 
\begin{eqnarray*}
(h _{n+1}^S)^T\left( x_{1},\ldots ,x_{n}\right)  &=& h_{n+1}^{S}
\left(
x_{1},\ldots ,x_{n+1}\right) -
\sum_{P\in \mathcal{I}_{n+1},P\neq
P_{0}}\prod_{r\in P} ( h _{\left| r\right| }^S )^T \left(
x_{r\left( 1\right) },\ldots ,x_{r\left( \left| r\right| \right) }\right)  \\
&=&
h _{n+1}^{S}\left(
x_{1},\ldots ,x_{n+1}\right) -\sum_{P\in \mathcal{I}_{n+1},P\neq
P_{0}}\prod_{r\in P} ( h _{\left| r\right| }^{T} ) ^{S}\left(
x_{r\left( 1\right) },\ldots ,x_{r\left( \left| r\right| \right) }\right)  \\
&=&h _{n+1}^{S}\left( x_{1},\ldots ,x_{n+1}\right) -
\left( \sum_{P\in 
\mathcal{I}_{n+1},P\neq P_{0}}\prod_{r\in P}h _{\left| r\right|
}^{T}\right) ^{S}\left( x_{1},\ldots ,x_{n+1}\right)  \\
&=&( h _{n+1}^{T}) ^{S}\left( x_{1},\ldots ,x_{n+1}\right), 
\end{eqnarray*}
where in the first line we have used the definition of the 
truncated $n$-point functions, in the second line we have used 
the induction hypothesis, and where in the third line 
we have applied formula \eqref{cmp}, applied to the 
hierarchy $\{h_1^T, h_2^T, \dots, h_n^T, \dots\}$.
This completes the induction. \qed

\subsection{Proof of Lemma \ref{lem0}}

Let $\Gamma$ be a closed conic subset of $T^*M^n \backslash (V^n_+ \cup V_-^+)$ and
let $\{t_\kappa\}$ be a sequence of smooth functions on $M^n$ 
, whose support 
is contained in some compact subset of $M^n$ for all $\kappa$, 
and which converges to some $t$ in the sense of 
$\cD'_\Gamma$. 
Then, the sequences $A_\kappa = \,\, : \varphi^{\otimes n}(t_\kappa) :_\omega$
by definition converges in $\mathcal W$. Therefore, since 
the state $\psi$ is assumed to be continuous on $\mathcal W$, 
$\psi(A_\kappa) = \Psi_n(t_\kappa)$ is a convergent 
sequence for $\kappa \to \infty$. 
We need to show that this implies that 
${\rm WF}(\Psi_n) \subset V_+^n \cup V_-^n$. 
This immediately follows from the following general result. 

\begin{lem}
\label{lem7}
Let $u \in \cD'(\mr^n)$ and let $\Gamma$ be a closed conic set in $\mr^n \times (\mr^n \backslash
\{0\})$. Assume that $u$ has the following property. For every sequence of smooth functions
$\{f_\kappa\}$ such that $f_\kappa \to f$ in $\cD'_\Gamma(\mr^n)$ and such that ${\rm supp}(f_\kappa)
\subset K$, with $K$ a compact subset of $\mr^n$, we have that 
$\{u(f_\kappa)\}$ is a 
convergent sequence. Then $\{0\} \notin {\rm WF}(u) + \Gamma$. 
\end{lem}

\paragraph{Proof.}

Let $A$ be a properly supported pseudo differential operator and 
with $\mu{\rm supp}(A) \subset \Gamma$.
Then, if $\{f_\kappa\}$ is any sequence 
of distributions on $\mr^n$ converging weakly to some $f \in \cD'(\mr^n)$ in the sense 
of distributions, it follows that $A f_\kappa \to Af$ in 
the sense of $\cD'_\Gamma(\mr^n)$ and that ${\rm supp}(Af_\kappa) \subset K'$, 
where $K'$ is some compact subset of $\mr^n$. Therefore, the sequence 
$u(Af_\kappa) = A^t u (f_\kappa)$ is convergent, by the assumption of the 
lemma, where $A^t$ is 
the formal adjoint of the pseudo differential operator $A$. We claim that it follows from 
this that $A^t u$ is in fact smooth. Assuming for the moment that this has been shown to 
be true, we get from the characterization \cite{h} of the wave front set that
\begin{equation*}
{\rm WF}(u) = \bigcap_{Bu \in C^\infty} {\rm char}(B) \subset 
\bigcap_{\mu{\rm supp}(A) \subset \Gamma} {\rm char}(A^t) = 
\bigcap_{\mu{\rm supp}(A) \subset \Gamma} -{\rm char}(A) = T^* \mr^n \backslash (-\Gamma),  
\end{equation*}
because the set of properly  supported pseudo differential operators $A$ with 
$\mu{\rm supp}(A) \subset \Gamma$ contains elements whose characteristic, 
${\rm char}(A)$, is contained in an arbitrarily small conic neighborhood of
$T^*\mr^n \backslash \Gamma$. Since the set $T^* \mr^n \backslash (-\Gamma)$ contains 
no element $(x, k)$ such that $k + k' = 0$ for some $(x, k') \in \Gamma$, this then 
proves the lemma. 

It thus remains to be shown that the compactly supported distribution 
$v = A^t u$ is smooth. This would immediately follow if the 
Fourier transform, $\widehat v(k)$ was rapidly decaying. Let us assume on the 
contrary that $\widehat v (k)$ is not rapidly decaying. We will show that this 
assumption is in contradiction with the fact that the sequence $v(f_\kappa)$ is 
converging for any weakly convergent sequence of distributions $f_\kappa$. So 
let us assume that $\widehat v(k)$ is not rapidly decaying. 
Then there exists an $N$, a positive constant
$C$, and a sequence of $k_j \in \mr^n$ with $|k_j| > j$ such that 
\begin{equation}
|\widehat v(k_j)| \ge C(1 + |k_j|)^{-N} \quad \text{for all $j$.}
\end{equation}
(We adopt from now on the convention that all strictly positive constants
appearing in the various inequalities are denoted by the same letter, $C$, 
irrespective of their numerical value.)
Let now $e \in \mr^n$ with $|e| = 1$ and let $t$ be a real number with 
$|t| \le r_j$, where $r_j > 0$. Then we have 
\begin{eqnarray}
|\widehat v(k_j + te) - \widehat v(k_j)| &\le& \int_0^{r_j}
\left|e \cdot \partial \widehat {v} (k_j + te)\right| \,dt\nonumber\\
&\le& \int_0^{r_j} C(1 + |k_j| + |t|)^{M} \,dt\nonumber\\
&\le& Cr_j(1 + |k_j| + |r_j|)^{M} 
\end{eqnarray}
for some $M$. This is easily seen to imply
\begin{equation}
|\widehat v(l)| \ge C(1 + |k_j|)^{-N} \quad \text{for all $j$}
\end{equation}
and all $l \in B_{r_j}(k_j)$, where $r_j = C(1 + |k_j|)^{-N-M}$ and 
$C>0$ is some sufficiently small constant. Let now $\rho_j$ be a function of 
compact support which is equal to one on $B_{r_j}(k_j)$ and which is 
zero outside $B_{2r_j}(k_j)$ and which satisfies $0 \le \rho_j(l) \le 1$
for all $l \in \mr^n$. Then, for any natural number $\kappa$ we define
a smooth function $f_\kappa$ on $\mr^n$ by 
\begin{equation}
\widehat f_\kappa(l) =
\overline{\widehat v(-l)} \sum_{j=1}^\kappa (1+|k_j|)^{L} \rho_j(-l),  
\end{equation} 
where $L$ is a constant to be chosen below.
Note that, since 
\begin{equation}
\left|(1 + |k_j|)^{L} \rho_j(l) \right| \le C(1 + |l|)^{L}
\quad \text{for all $j$,}
\end{equation}
the functions $f_\kappa$ converge weakly to some distribution $f$. 
By what we have already said above, this means that the sequence 
$v(f_\kappa)$ should converge as $\kappa \to \infty$. We will now 
show that this sequence in fact manifestly diverges, leading thus 
to the desired contradiction. Inserting the definition of $f_\kappa$, 
we get  
\begin{eqnarray}
v(f_\kappa) &=& \int_{\mr^n} \widehat v(l) \widehat f_\kappa(-l) \, d^n l \nonumber\\ 
&=  & \sum_{j=1}^\kappa \int_{\mr^n} |\widehat v(l)|^2 (1 + |k_j|)^{L} \rho_j(l)\, 
d^n l \nonumber\\
&\ge& C\sum_{j=1}^\kappa \int_{B_{r_j}(k_j)} (1 + |k_j|)^{-2N+L} \, d^n l \nonumber\\
&\ge& C\sum_{j=1}^\kappa (1 + |k_j|)^{-2N+L}r_j^n \nonumber\\
&\ge& C\sum_{j=1}^\kappa (1 + |k_j|)^{-2N-n(N+M)+L}.
\end{eqnarray}
If $L$ is now chosen large enough, then the right side of 
the above inequality diverges for $\kappa \to \infty$, since $k_j > j$. 
\qed
\qed


\begin{thebibliography}{99}

\bibitem{bphz}
See for example the lectures given at the 1975 Erice 
summer school, collected in: 
``Renormalization Theory,'' G. Velo and A. S. Wightman, eds., 
NATO ASI Series C {\bf 23}, Reidel, Dodrecht, 1976

\bibitem{bu}
T. S. Bunch: ``BPHZ renormalization of $\lambda \Phi^4$ field theory 
in curved space-times,'' Ann. of Phys. 131, 118 (1981) 

\bibitem{p}
T. S. Bunch, P. Panangaden and L. Parker: ``On 
renormalization of $\lambda\Phi^4$ in curved space-time I,'' 
J. Phys. A: Math. Gen. {\bf 13}, 901-918 (1980), 
``On renormalization of $\lambda\Phi^4$ in curved space-time II,'' 
J. Phys. A: Math. Gen. {\bf 13}, 919-932 (1980) 

\bibitem{bfk}
R. Brunetti, K. Fredenhagen and M. K\"ohler: ``The microlocal spectrum 
condition and Wick polynomials on curved spacetimes,'' Commun. Math.
Phys. {\bf 180}, 633-652 (1996) 

\bibitem{bf}
R. Brunetti and K. Fredenhagen: ``Microlocal Analysis and 
Interacting Quantum Field Theories: 
Renormalization on physical backgrounds,''  
Commun. Math. Phys. {\bf 208}, 623-661 (2000) 

\bibitem{hw}
S.~Hollands and R.~M.~Wald:
``Local Wick polynomials and time ordered products of quantum fields in  curved spacetime,'' Commun. Math. Phys., in print, [gr-qc/0103074].

\bibitem{h}
L. H\"ormander: ``The Analysis of Linear Partial Differential Operators
I--IV'', Springer-Verlag, Berlin 1985

\bibitem{hw1}
S.~Hollands and R.~M.~Wald: work in progress.

\bibitem{wald}
R. M. Wald: ``Quantum Field Theory in Curved Spacetime and Black Hole
Thermodynamics,'' The University of Chicago Press, Chicago 1994 

\bibitem{kw}
B. S. Kay and R. M. Wald: ``Theorems on the uniqueness and thermal
properties of stationary, nonsingular, quasifree states on spacetimes
with a bifurcate Killing horizon,'' Phys. Rep. {\bf 207}, 49 (1991)

\bibitem{r}
M. J. Radzikowski: ``Micro-Local Approach to the Hadamard condition in
QFT on Curved Space-Time,'' Commun. Math. Phys. {\bf 179}, 529-553 (1996) 

\bibitem{fd}
M.~D\"utsch and K.~Fredenhagen:
``Algebraic quantum field theory, perturbation theory, 
and the loop  expansion,''
[hep-th/0001129]; 
``Perturbative algebraic field theory, and deformation quantization,''
[hep-th/0101079].

\bibitem{lr}
C. L\"uders and J. Roberts: ``Local quasiequivalence and adiabatic 
vacuum states,'' Commun. Math. Phys. {\bf 134}, 29 (1990).

\bibitem{ho}
S.~Hollands:
``The Hadamard condition for Dirac fields and adiabatic 
states on  Robertson-Walker spacetimes,''
Commun.\ Math.\ Phys.\  {\bf 216}, 635 (2001)
[gr-qc/9906076].

\bibitem{j}
W. Junker: in preparation.

\bibitem{v}
R.~Verch:
``Local definiteness, primarity and quasiequivalence of quasifree 
Hadamard quantum states in curved space-time,''
Commun.\ Math.\ Phys.\  {\bf 160}, 507 (1994).

\bibitem{hdp}
Handbuch der Physik, Band XII, ``Thermodynamik der Gase,'' 
Springer-Verlag 1958

\end{thebibliography}
\end{document}